\documentclass[12pt]{article}%
\usepackage{times}%
\usepackage{amsmath}%
\usepackage{amsfonts}%
\usepackage{amssymb}%
\usepackage{graphicx}

\begin{document}

\title{\textbf{Weak Quantum Theory }\\\textbf{and the Emergence of Time}}
\author{\textbf{Hartmann R\"{o}mer}$\,\medskip\thanks{email:
hartmann.roemer@physik.uni-freiburg.de}$\\{\normalsize $\,$ }Fakult\"{a}t f\"{u}r Physik der Universit\"{a}t Freiburg\\Hermann-Herder-Str.~3\\D-79104 Freiburg, Germany}
\maketitle

\begin{abstract}
We present a scenario, how time could emerge in the framework of Weak Quantum
Theory. In a process, similar to the emergence of time in quantum cosmology,
time arises after an epistemic split of the unus mundus as a quality of the
individual conscious mind. Synchronization with matter and other mental
systems is achieved by entanglement correlations. In the course of its
operationalization, time loses its original quality of A-time and the B-time
of physics as measured by clocks will appear.

\end{abstract}

\newpage

\section{Introduction}

The task of this study is the establishment and description of a scenario for
the emergence of time in the framework of Weak Quantum Theory \cite{wqt}, a
generalization of quantum theory applicable beyond the ordinary domain of
physics but containing essential quantum theoretical features like
complementarity and entanglement.

The mysterious origin and nature of time have ever since been a permanent
subject of human thinking and philosophy \cite{ttn}. Later, also physics and,
more recently, brain physiology and neuroscience \cite{poeppel} have
contributed to these questions. Time is given to us in two very different
forms: first as internal time, as an immediate mode of our personal existence
and secondly as external time, the kind of entity which appears in physics and
is measured by clocks. Employing a distinction introduced by McTaggart
\cite{mctaggart}, internal time can be characterized as \textit{A-time}: there
is an essential quality of ''nowness'' which distinguishes presence from past
and future. Presence is continuously moving into the future and thereby
turning into past. A-time may also admit additional qualities: good and
favorable or bad and unfavorable for certain tasks. The Greek notion of
$\kappa\alpha\iota\rho\acute{o}\varsigma$ is an example for a quality
associated to A-time. A-time is also called ''tempus'' as opposed to ''time'',
because it underlies the tempora of the verb in many human languages. On the
other hand,\textit{\ B-time} is the time of physics. All points of B-time are
equivalent and void of any additional qualities, they are just points on a
linear scale, the only and fundamental distinction being a (partial) ordering
in the sense of \ ''earlier'' and ''later''. Even this directedness of B-time
is absent in physics, if time inversion symmetry is assumed to hold.

The questions about internal and external time cannot be addressed without
reference to the problem of the relationship between mind and matter. Here, we
cannot enter into a deeper discussion of this complicated complex of problems,
which has a long history and is presently a subject of intensive discussion
and research \cite{mattermind}. We can only classify the positions which are
logically possible in order to provide a coordinate system in which we can
locate our own standpoint.

The first and principle distinction is between \textit{dualistic} and
\textit{monistic} conceptions of mind and matter.

In modern philosophy, Descartes \cite{descartes} is usually considered to be
the first and most prominent proponent of dualism. Under the terms ''res
cogitans'' and ''res extensa'', mind and matter are fundamentally different
substances of different ontological status. The main problem of all dualistic
theories is the explanation of mutual causation or, more generally, of
correlations between mind and matter. After all, mind and matter go together
and both of them take part in most events happening with us in our world.
Several solutions have been proposed among which we only mention Descartes'
theory of causation, the occasionalism of Malbranche \cite{malbranche} and, in
particular, Leibniz' \cite{leibniz} notion of prestabilized harmony. According
to Leibniz, matter and mind always go in parallel, not because of any
interaction between them but because they are perfectly synchronized by their
divine creator. Many people have noticed the striking similarity between such
a prestabilized harmony and the interactionless correlations appearing in
quantum systems in entangled states.

It is fair to say that at present, partly because of the difficulties
mentioned, dualistic approaches have largely fallen out of favour.

Monistic theories of mind and matter deny the existence of two separate
substances for them. According to the degree of priority attributed to either
matter or mind, they can be classified in a threefold way.

\begin{itemize}
\item \textit{Matter over mind} theories consider some form of matter to be
the only fundamental substance of the world. There are large differences
between the various concepts of matter in such theories. If mind is at all
admitted as a decent object of investigation, it is conceived as an
epiphenomenon, a feature of the ''\"{U}berbau'' or as an emergent feature of
matter. Again, there is a plethora of different conceptions of emergence. The
majority of working scientists, biologists and neurophysiologists even more
than physicists, seem to favour some version of a matter over mind theory,
which, in addition, appears to be supported by the impressive success of
modern science and fits in very well with the widespread materialistic view of
the world.

\item \textit{Mind over matter} theories are adopted in a rather diffuse way
by many esoteric circles. An intellectually viable example of this conception
is the philosophy of Hegel \cite{hegel}, for whom the substance of the world
is of genuinely spiritual nature such that events in the material world are
manifestations of the dynamic and dialectic self-reflection of this universal spirit.

\item \textit{Neutral monistic} theories consider matter and mind to be
different manifestations of equal right of one and the same substance, which
in itself is neither matter nor mind. This is the point of view we shall
adopt. It is presently gaining ground among professionals \cite{chalmers} of
the mind and matter problem. It was clearly formulated by Spinoza
\cite{spinoza}, for whom, out of a possible infinity of modi in which one and
the same universal substance could manifest itself, mind and matter are just
those two modi which are accessible for human beings. More recently, neutral
monism has been advocated by C.G. Jung. He started out from his theory of the
collective unconscious, an extension of the individual mind into a
transpersonal collective domain of psychic character, regulated by general
abstract but emotionally loaded patterns which he called \textit{archetypes.}
Later and partly under the influence of W. Pauli \cite{paulijung},
\cite{jungpaulibriefe}, the archetypes turned into even more abstract
regulating principles within the domain of the \textit{unus mundus, }which is
imagined to be neutral with respect to the distinction between mind and
matter. Synchronistic phenomena like the so-called meaningful coincidences
could thus be described as partly physical and partly psychic manifestation of
archetypal configurations. It was in particular W. Pauli, one of the fathers
of quantum theory, who compared this structure with quantum theory and
conceived the distinction of matter and mind as a kind of symmetry breaking in
the \textit{unus mundus}. Material and mental descriptions of the \textit{unus
mundus} could thus be considered as complementary in the sense of quantum
theory. In the same way the causal order in the physical world and the order
of sense and meaning were interpreted as complementary.
\end{itemize}

The above-mentioned threefold distinction has, of course a bearing on the
problem of the relationship between internal and external time. For instance,
a materialist would probably postulate a priority of the physical B-type time
and consider internal A-time to be a derived notion. However, it has turned
out to be extremely difficult to derive the directedness of the time in
thermodynamics as well as the directedness of internal time from a time
symmetric physical background. Indeed, none of the proposed derivations of the
''arrow of time'' is completely satisfactory. On closer inspection, almost all
of these derivations, with the possible exception of the cosmological time
arrow, either explicitly or silently take recourse to the psychological arrow
of internal time. Even more difficult is the derivation of the unique and
characteristic quality of ''now'' in internal time, in fact, to such an extent
that this problem is often ignored, defined away or declared meaningless.

Our own approach presented in this study starts out from a neutral monistic
conception of mind and matter. We shall locate the origin of time in the
personal consciousness assuming that time is essentially and intimately
related to our form of existence as conscious individual beings. Supporting
evidence for this assumption comes from the common observation that the
unconscious dimension does not seem to know about time. Already in dreams the
dimension of time starts fading away and the deeper parts of unconscious and,
even more so, the collective unconscious are entirely timeless. Also C. G.
Jung's \textit{unus mundus} is explicitly assumed to be timeless
\cite{paulijung}. There is a long tradition in philosophy relating time to our
form of existence. For Augustinus \cite{augustinus}, A-time is the mode and
limitation of the finite rather than infinite existence of human beings. For
Immanuel Kant \cite{kant} time is similar to Newton's B-time. He considers
time to be the form of the interior sense of humans, prior to and a
prerequisite for any act of cognition. Also in the 20th century philosophy of
existence \cite{husserl}, \cite{heidegger} A-time is tied to human existence
as an essential determining feature. There are, of course, alternatives to our
approach. For instance H. Primas \cite{primas2003}, in a remarkable study
about the origin of time, associates a time of B-type primarily to a
collective mental dimension of the world.

Our strategy in attacking the problem of the origin of time is to apply Weak
Quantum Theory to a primarily undivided \textit{unus mundus}. The main theses
we shall try to develop are \cite{offenburg}

\begin{itemize}
\item The \textit{unus mundus} is timeless and neutral with respect to the
distinction of mind and matter. This distinction only arises after an
''epistemic splitting'' of the \textit{unus mundus} by separating a
''conscious observer'' from the rest of the world. Observables pertaining to
mind and matter aspects of the \textit{unus mundus }are in general
complementary\textit{. }Such a splitting is the prerequisite for and
inevitably connected with any act of cognition in the most general sense that
someone arrives at knowledge or information about something. It is only after
this epistemic split that time can arise. One should notice that also animals
can learn about their surroundings and have some sense of time. This means
that full human consciousness of the ''conscious observer'' is not required
for the epistemic split and the emergence of time. Primarily, time emerges as
A-time, related to the conscious observer. The process of emergence shows a
formal analogy with the arrival of time in the Wheeler-de Witt equation
\cite{wheeler} of quantum cosmology where the quantum state of the universe
allows for the interpretation of certain observables as time observables.

\item The transfer of time to material systems and the synchronization with
other observers and material subsystems are effected by entanglement
correlations due to the state of the \textit{unus mundus.}

\item Physical B-time arises by a complicated process of redefinition, gauging
and operationalization certainly requiring full human consciousness. In the
course of this process, time loses most of its qualities and may eventually
disappear by ''deconstruction''.

The material of this work is organized in the following way:
\end{itemize}

In Chapter 2 we provide the minimum of Weak Quantum Theory necessary for
following our arguments.

Chapter 3 is devoted to the somewhat problematic notion of the set of
observables and the state of the universe. The crucial role of the epistemic
split and, as a consequence, the observer dependence of the set of observables
are pointed out. In addition, we describe, how physical quantum theory can be
embedded into Weak Quantum Theory.

Chapter 4 starts with a description of a toy model for the Wheeler-de Witt
equation. It illustrates how time can arise as a property of the quantum state
in an initially timeless situation. Subsequently, we briefly describe how time
can be introduced in cosmology by a solution of the Wheeler-de Witt equation.

In Chapter 5 a partially analogous scenario for the emergence of time in Weak
Quantum Theory is worked out.

Chapter 6, which is more than an appendix will contain additional remarks,
questions and speculations.

In spite of mutual independence, there will be some overlap between our work
and the ingenious study of H. Primas \cite{primas2003}, in particular
concerning the importance of symmetry breaking in the \textit{unus mundus }and
the function of entanglement correlations. Similarities and differences of our
approaches will be mentioned along with our presentation. To the educated
reader, many of our ideas will not be unfamiliar, a situation to be expected
for such an old subject under vivid actual discussion. We still hope that she
or he will appreciate our kind of synthesis as well as quite a few novel features.

\bigskip

\section{A Sketch of Weak Quantum Theory}

Weak Quantum Theory is a generalization of quantum theory devised to be
applicable beyond the range of ordinary physical systems. It was obtained
starting out from the algebraic formulation of quantum theory and relaxing all
those axioms which seem to be special to the physical world. The remaining
more general structure is still rich enough to be able to describe quantum
like phenomena like complementarity \cite{wal} and entanglement in a much more
general setting. Here, we give a short sketch of the structure of Weak Quantum
Theory just sufficient to make the presentation in this work reasonably self
sustained. For details as well as for several applications of Weak Quantum
theory we refer to the original publications \cite{wqt}, \cite{bistable}.

In Weak Quantum Theory, the fundamental notions of \textit{system, state} and
\textit{observable} are taken over from ordinary quantum theory:

\begin{itemize}
\item A system $\Sigma$ is any part of reality in the most general sense,
which can, at least in principle, be isolated from the rest of the world and
be the subject of an investigation.

\item A system is assumed to have the capacity to reside in different states.
The notion of state also has an epistemic side, reflecting the degree of
knowledge of an observer about the system. Unlike in ordinary quantum
mechanics, the set $\mathcal{Z}$ of states is not assumed to have an
underlying linear Hilbert space structure.

\item An observable $A$ of a system $\Sigma$ is any feature of $\Sigma$ which
can be investigated in a (more or less) meaningful way. Let $\mathcal{A}$
denote the set of observables. Just like in ordinary quantum mechanics,
observables $A$ in $\mathcal{A}$ can be identified with functions on the set
of states: Any observable $A\in\mathcal{A}$ associates to every state
$z\in\mathcal{Z}$ another state $A(z)\in\mathcal{Z}$. As functions on the set
of states, observables $A$ and $B$ can be composed by applying $A$ after $B$.
The composed map $AB$ is also assumed to be an observable. Observables $A$ and
$B$ are called \textit{compatible }or\textit{\ commensurable }if they commute,
i.e. if $AB=BA$. Noncommuting observables with $AB\neq BA$ are called
\textit{complementary} or \textit{incompatible}. In ordinary quantum theory,
observables can also added, multiplied by complex numbers and conjugated, and
the set of observables is endowed with a rich structure called $C^{\ast}%
$-algebra structure. In Weak Quantum Theory, observables can only be
multiplied by the above composition. This gives the set of observables a much
simpler so-called \textit{semigroup structure.}

In ref \cite{wqt}, Weak Quantum Theory is characterized by a list of axioms.
Here, we only give the most important properties:

\item To every observable $A\in\mathcal{A}$ there is an associated set
$specA$, which is called its \textit{spectrum. }The set $specA$ is just the
set of different outcomes or results of the investigation (''measurement'')
corresponding to the observable $A$.

\item \textit{Propositions} are special observables $P$ with $PP=P$ and
$specP\subset\left\{  yes,no\right\}  $. They simply correspond to yes-no
questions about the system $\Sigma$. For every proposition $\ P$ there is a
negation $\bar{P}$ compatible with $P$. For compatible propositions
$P_{1\text{ }}$ and $P_{2}$ there exists a conjunction $P_{1}\wedge
P_{2}=P_{1}P_{2}$ and an adjunction $P_{1}\vee P_{2}=\overline{\overline
{P_{1}}\wedge\overline{P_{2}}}$. The laws of ordinary proposition logic are
assumed to hold for compatible propositions.

\item If $z$ is a state and $P$ is a proposition with $P(z)\neq0$, then $P(z)$
is a state for which $P$ is true with certainty. This emphasizes the
constructive nature of measurement as preparation and verification.

\item The following property generalizes the spectral property of observables
in ordinary quantum theory. To every observable $A$ and every element
$\alpha\in specA$ there belongs a proposition $A_{\alpha}$ which is just the
proposition that $\alpha$ is the outcome of a measurement of $A$. Then%

\begin{equation}
A_{\alpha}A_{\beta}=A_{\beta}A_{\alpha}=0\ \;\text{for }\alpha\neq\beta,\quad
AA_{\alpha}=A_{\alpha}A,\quad\bigvee_{\alpha\in specA}A_{\alpha}%
={\mathchoice{\rm1\mskip-4mu l} {\rm1\mskip-4mu l} {\rm1\mskip-4.5mu l}
{\rm1\mskip-5mu l}}\label{spec}%
\end{equation}

where $0$ and ${\mathchoice{\rm1\mskip-4mu l} {\rm1\mskip-4mu l} {\rm
1\mskip-4.5mu l} {\rm1\mskip-5mu l}}$ are just the trivial propositions which
are never and always true respectively.
\end{itemize}

We already mentioned that Weak Quantum Theory is rich enough to encompass the
notions of complementarity and entanglement. For complementary observables
$\ A$ and $B$ with $AB\neq BA$, the order of their measurement matters, and,
just like in ordinary quantum mechanics, they will not, in general, possess
states in which both of them have a well defined value with certainty.

Entanglement arises if global observables pertaining to all of a system
$\Sigma$ are complementary to local observables pertaining to parts of
$\Sigma$. In an entangled state, for instance in a state in which a global
observable has a well defined value, there are typical interactionless
entanglement correlations between the results of measurements of local
observables. In ordinary quantum theory, it can be proved that entanglement
cannot be used for signal transmission or causal intervention. In Weak Quantum
Theory, it may be wise to postulate this feature \footnote{I am grateful to
Walter von Lucadou for pointing this out to me} as an additional axiom
\cite{offenburg} supplementing the axioms of ref \cite{wqt}.

Notice, that Weak Quantum Theory, at least in its minimal version presented
here, does not associate quantified probabilities to the outcomes of a
measurement of an observable $A$. This is related to the absence of a Hilbert
state structure of the set $\mathcal{Z}$ of states. Moreover, the notion of
time is completely absent in the general formulation of Weak Quantum Theory.

Planck's constant $h$ which controls the degree of noncommutativity in
ordinary quantum theory, does not enter into Weak Quantum Theory.

At this place, we should like to mention another possible enrichment
\cite{offenburg} of the axioms of ref \cite{wqt}, to which we shall return at
the end of this study. One could admit a more general kind of observables
without an associated spectrum, for which the name \textit{preobservables
}might be appropriate. Preobservables are meant to correspond to an
expectationless precategorical state of attention of the observer. Only after
the establishment of a horizon of expectations as a result of additional
experience, it may become possible to associate a spectrum to them and to turn
them into full ordinary observables. We hope to come back to this issue in a
separate publication.

\section{Observables and Epistemic Splitting}

Weak Quantum Theory is a very general theory meant to be applicable to all
kinds of systems $\Sigma$, which can be singled out for investigation from the
rest of the world. What we have in mind here, is an application of Weak
Quantum Theory to the totality of the \textit{unus mundus.}This is not an
unproblematic enterprise. The same problem arises in quantum cosmology, where
ordinary quantum mechanics is applied to the whole cosmos: the very name of an
''observable'' betrays, that the existence of an observer outside the observed
system is presupposed and that both ordinary and Weak Quantum Theory primarily
apply to the description of systems as seen from an outside observer. In what
way does it make sense to talk about the wave function of the universe or the
state of the \textit{unus mundus}?

First of all, it is always possible to enlarge a system $\Sigma_{1}$ by
inclusion of parts $\Sigma_{2}$ previously outside $\Sigma_{1}$. For example,
one may include the observer of a system $\Sigma$ into a larger system and
study the interaction of $\Sigma$ with its observer in the enlarged system
(possibly as observed by a ''superobserver'').

In ordinary quantum theory, there exists a canonical tensor product
construction for the Hilbert space of states and the algebra of observables of
a composite system from its components. This is not at our disposal in Weak
Quantum Theory, but one can at least say \cite{wqt} that the state space and
the semigroup of observables of a composed system will contain the Cartesian
product of the state spaces and observable semigroups of its components:%

\begin{align}
\mathcal{A}  & \supset\mathcal{A}_{1}\times\mathcal{A}_{2},\;\ Z\supset
\mathcal{Z}_{1}\times\mathcal{Z}_{2},\label{composition}\\
\mathcal{A}_{1}(\mathcal{Z}_{1})  & \subset\mathcal{Z}_{1},\ \;\mathcal{A}%
_{2}(\mathcal{Z}_{2})\subset\mathcal{Z}_{2}.
\end{align}

At least as important as the enlargement of systems is the possibility of
analyzing systems by identifying subsystems in them, whose mutual relationship
can be investigated. This act of decomposition into subsystems is a
constitutive mental act. There are infinitely many ways to decompose a system
into subsystems, and the kind of decomposition is not dictated by the system
itself. Rather, the system as such remains unchanged after decomposition. In
this sense, the decomposition is a purely mental act. On the other hand, it is
fair to say, that it is only by decomposition, that the subsystems come into
being, which underlines the creative status of decomposition. Mahler
\cite{mahler} strongly points out this double importance of decomposition and
takes it as the point at which consciousness can intervene in our world. In
ordinary quantum mechanics, decomposition corresponds to a tensor product
decomposition of the Hilbert space of states and the algebra of observables:%

\begin{equation}
\mathcal{H=H}_{1}\mathcal{\otimes H}_{2},\;\;\mathcal{A=A}_{1}\mathcal{\otimes
A}_{2}\label{decomp}%
\end{equation}
and in Weak Quantum Theory subsemigroups and subsets of states have to be
identified fulfilling eq \ref{composition}. The decomposition of a system into
subsystems can be considered as a symmetry breaking, because it introduces
distinctions which are not dictated by the system itself.

In view of the twofold possibility of composition and decomposition or of
synthesis and analysis, talking about the universe or the \textit{unus mundus
}as a system appears to be a reasonable extrapolation. This kind of
extrapolation is, for instance, employed in quantum cosmology, where ordinary
quantum theory is applied to the universe as a whole. In Weak Quantum Theory
where no probabilities are attributed to measurements, the problem may even be
alleviated somewhat, because an ensemble interpretation seems to be less
mandatory than for ordinary quantum theory.

The first and most important act of decomposition is the epistemic splitting,
the inevitable starting point of any act of cognition, whereby a observer is
set apart from what he/she observes. We already mentioned that the notion of
an observable presupposes an epistemic split. Moreover the epistemic split is
unresolvably connected to the appearance of consciousness in however
rudimentary form. What is required is that some entity is set apart from the
rest of the world in maintaining itself, gaining information about its
environment and reacting to it. Higher levels of consciousness also involve a
capacity to form a self representation in a self model as described in detail
by Th. Metzinger \cite{metzinger}. Observations in the technical sense will
require such higher states of consciousness.

Weak Quantum Theory has to face the problem to explain how the material or
physical world can be embedded into a supposedly larger system possessing also
nonmaterial features. This can actually be achieved in the following way:

Inside the large semigroup of observables there is a subsemigroup of material observables:%

\begin{equation}
\mathcal{A}_{matter}\subset\mathcal{A}%
\end{equation}

Now, $\mathcal{A}_{matter}$ has the richer structure of a $C^{\ast}-$algebra.
A state $z\in\mathcal{Z}$ gives rise to a positive linear complex valued
expectation value functional $E_{z}$ defined on $\mathcal{A}_{matter}$:%

\begin{align}
E_{z}(\alpha A+\beta B)  & =\alpha E_{z}(A)+\beta E_{z}(B)\\
E_{z}(A^{\ast}A)  & \geq0
\end{align}

for complex $\alpha,\beta$ and $A,B$ in $\mathcal{A}_{matter}$. For
observables $A\in\mathcal{A}_{matter}$, the spectrum $specA$ should be
contained in the set of complex numbers.

This establishes the ordinary probability interpretation for quantum theory in
the material world. Planck's constant $h$ will play its role in $\mathcal{A}%
_{matter}$. Two states $z$ and $z^{\prime}$ are called \textit{physically
equivalent,} if their associated expectation value functionals coincide.%

\begin{equation}
z\sim z^{\prime}\Leftrightarrow E_{z}(A)=E_{z^{\prime}}(A)\;\;\text{for all
}A\in\mathcal{A}_{matter}%
\end{equation}

The resulting equivalence classes should be called \textit{physical states.}
Matter observables $A\in\mathcal{A}_{matter}$ will transform physical states
into physical states. This is not expected to be true for other observables in
$\mathcal{A}$. Starting from any physical state, a physical Hilbert space can
be obtained by the GNS construction \cite{haag}. As a linear operator on a
Hilbert space and also as an element of a $C^{\ast}-$algebra, every observable
$A\in\mathcal{A}_{matter}$ will have a spectrum $SPEC\;A$ and we shall have
$SPEC\;A=specA$.

Knowing $\mathcal{A}_{matter}$ is is natural to ask about its
\textit{commutant} $\mathcal{A}_{matter}^{\prime}$ which consists of all
observables of the weak quantum theoretical system commuting with all material observables:%

\begin{equation}
\mathcal{A}_{matter}^{\prime}=\left\{  B\in\mathcal{A\mid}BA=AB\text{ for all
}A\in\mathcal{A}_{matter}\right\}
\end{equation}

Primas \cite{primas2003}, in the framework of ordinary quantum theory,
essentially identifies $\mathcal{A}_{matter}^{\prime}$ with the subalgebra of
mental observables and assumes a decomposition of the Hilbert space and the
observable algebra of the \textit{unus mundus }of the kind%

\begin{equation}
\mathcal{H=H}_{matter}\mathcal{\otimes H}_{\text{\textit{mind}}}%
,\;\;\mathcal{A=A}_{matter}\mathcal{\otimes A}_{\text{\textit{mind}}%
}\label{primdec}%
\end{equation}

This means that matter and mind observables always commute.

We prefer a complementary relationship between matter and mind, in accordance
with the intention of W. Pauli and C. G. Jung \cite{paulijung},
\cite{jungpaulibriefe}. For instance, under the headings of ''brain'' and
''mind'' one and the same system can be investigated in two very different
ways, either physiologically with the methods of physical observation and
experimentation or psychologically by introspection, redirection of self
attentiveness and reporting about them. These two approaches will use
complementary ''matter'' and ''mind'' observables respectively. So, for us,%

\begin{equation}
\mathcal{A}_{\text{\textit{mind}}}\cap(\mathcal{A\setminus A}_{matter}%
^{\prime})\neq\emptyset\label{matmind}%
\end{equation}

\section{The Wheeler-de Witt Equation and Cosmological Time}

In this section, we want to describe, how time can be introduced in cosmology
in a primarily timeless framework by means of a solution of the Wheeler-de
Witt equation. It is our intention to generalize this scheme to Weak Quantum
Theory, which will be done in the subsequent section.

The essentials of the principle can best be understood from a very simple toy model.

Consider a system in ordinary quantum theory, whose algebra of observables is
generated by two observables $X$ and $Y$ together with their conjugates
$P_{X}$ and $P_{Y}$ \ The fundamental commutation relations are just the
commutation relations for position and momentum of a point particle in two
dimensional space:%

\begin{align}
\left[  X,Y\right]   & =\left[  P_{X},P_{Y}\right]  =\left[  X,P_{Y}\right]
=\left[  Y,P_{X}\right]  =0\\
\left[  X,P_{X}\right]   & =\left[  Y,P_{Y}\right]  =i\frac{h}{2\pi
}{\mathchoice{\rm1\mskip-4mu l} {\rm1\mskip-4mu l} {\rm1\mskip-4.5mu l}
{\rm1\mskip-5mu l}}%
\end{align}

In a basis of simultaneous eigenstates $\left|  x,y\right\rangle $ of $X$ and
$Y$, state vectors of the system will be given by functions $\psi(x,y)$.
Assume now that the state function obeys an equation%
\begin{equation}
\left(  \frac{\partial^{2}}{\partial x^{2}}-\frac{\partial^{2}}{\partial
y^{2}}\right)  \psi(x,y)=0\label{schr}%
\end{equation}

In our simple example it is even possible to give the general solution of eq
\ref{schr} :%
\begin{equation}
\psi(x,y)=f(x-y)+g(x+y)\label{schrsol}%
\end{equation}

for arbitrary functions $f$ and $g$.

In general, the solution eq \ref{schrsol} does not factorize into a function
of $x$ and a function of $y$, although there are, of course, also special
factorizing solutions like
\begin{equation}
\psi(x,y)=\sin(kx)\sin(ky),\label{prod}%
\end{equation}
but generically the solution of eq \ref{schr} will not factorize but will be
entangled with respect to the observables $X$ and $Y$. Entangled solutions are
only representable as superpositions of factorizing solutions. Now, in
contrast to a factorizing solution like eq \ref{prod}, for an entangled
solution, the distribution of the values of $y$ will depend on the value of
$x$. In this sense, $x$ controls the knowledge of $y$. In the extreme case%
\begin{equation}
\left|  \psi\right\rangle =\int dx\;c(x)\left|  x,y(x)\right\rangle
\end{equation}
the value of $x$ even completely determines $y$, the other extreme is just
given by factorizing solutions like eq \ref{prod}.

This allows to interpret the controlling variable $x$ as a time variable,
whereby a factorizing solution would describe a time independent situation.
For a hyperbolic equation like \ref{schr}, the time like variable $x$ shares
another feature of time as it is normally understood in physics: Prescribing
the initial values for $x=0$:%
\begin{equation}
\psi(0,y)=a(y),\;\frac{\partial}{\partial x}\psi(0,y)=b(y)
\end{equation}
will completely fix the solution of the state equation \ref{schr} for all
values of the time $x$. This means that the hyperbolic character of eq
\ref{schr} leads to a deterministic time development with respect to $x.$

The Wheeler-de Witt equation \cite{wheeler} is an equation for the wave
function of the universe in quantum cosmology, which can be conceived as an
enormous upgrading of our toy model eq \ref{schr}. An infinity of pairs of
conjugate variables enters rather than just two, such that the variable $X$ is
replaced by the spacial metric $h_{ab}$ of the universe, and $Y$ corresponds
to an infinity of observables $\varphi$ pertaining to matter fields in the
universe. The derivatives in eq \ref{schr} are to be replaced by functional
derivatives with respect to $h_{ab}$ and $\varphi$. The wave function
$\psi(x,y)$ is replaced by a functional $\left|  \Psi\left[  h_{ab}%
,\varphi\right]  \right\rangle $ depending on the spacial metric and the
matter fields. The Wheeler-de Witt equation is a direct consequence of the
invariance of General Relativity Theory under arbitrary coordinate
transformations. It has a structure similar to eq \ref{schr}, which we write
down for the benefit of the reader with some familiarity in quantum field theory:%

\begin{equation}
\left\{  -\frac{1}{2m_{P}^{2}}G_{ab,cd}\frac{\delta^{2}}{\delta h_{ab}\delta
h_{cd}}-m_{P}^{2}\sqrt{h}R^{(3)}+H_{matter}\left[  h_{ab,}\varphi\right]
\right\}  \left|  \Psi\left[  h_{ab},\varphi\right]  \right\rangle
=0\label{wdw}%
\end{equation}

Here, $m_{P}$ is the so-called Planck mass, $h$ is the determinant of
$h_{ab\text{, }}R^{(3)}$ is the scalar curvature associated to $h_{ab}$ and
$G_{ab,cd}$ is a metric in the infinite dimensional ''superspace'' of spatial
metrics and given by%

\begin{equation}
G_{ab,cd}=\frac{1}{\sqrt{h}}\left(  h_{ac}h_{bd}+h_{ad}h_{bc}-h_{ab}%
h_{cd}\right) \label{metric}%
\end{equation}

$H_{matter}\left[  h_{ab,}\varphi\right]  $ is a term depending on the metric
and the matter fields, whose precise form depends on the model for the matter fields.

The Wheeler-de Witt equation \ref{wdw} does not contain any reference to time,
but, depending on the nature of its solution, a time variable can be
introduced in a way completely analogous to our toy model. The metric
$G_{ab,cd}$ is of hyperbolic character, and this opens up the possibility to
interpret one combination of the variables $h_{ab}$ as a time variable
monitoring a deterministic development of the other variables if the solution
of eq \ref{wdw} is not factorizing \cite{kiefer}.

Which variable precisely takes over the role of a time depends on the solution
of eq \ref{wdw}. Models have been constructed, whose solution corresponds to
an expanding universe, and in these models, it is the determinant function
$\sqrt{h}$, which takes over the role of a time variable. The quantity
$\sqrt{h}$ is directly related to the radius of the universe, which in an
expanding universe serves as a measure of time. The fact that time is normally
felt as a classical parameter rather than a quantum observable is explained by
a mechanism of \textit{decoherence} \cite{joos}. By the interaction with the
infinity of other degrees of freedom, the time operator $\sqrt{h}$ \ is
effectively measured continuously, and the state of the universe becomes
indistinguishable from an incoherent superposition of states with different
values of the time observable.

\section{Emergence of time in Weak Quantum Theory}

The core of the argument of the preceding section on the emergence of time in
quantum cosmology can be transferred to Weak Quantum Theory applied to the
state of the \textit{unus mundus. }A discussion in terms of Weak Quantum
Theory seems to be mandatory because we do not expect the formalism of full
ordinary quantum theory to be applicable at this level of generality.
Essentially, we locate time primarily in individual consciousness and assume
entanglement correlations to be the decisive mechanism for time
synchronization. The motivation of this approach are a neutral monistic
attitude towards the mind matter problem and the simple observations that time
is intimately related to our mode of existence as conscious individuals and
that our internal time shows a high degree of correlation with the internal
times of other individuals and with changes in the material world. This
suggests that the state of the \textit{unus mundus }is strongly
entangled.\ More precisely, our scenario is as follows:

\begin{enumerate}
\item We already mentioned in section 3 that individual consciousness, at
least at some low level, is intimately related to the epistemic split, which,
after all consists in the isolation of an observing subject from the rest of
the world. In addition, the distinction between matter and mind requires the
epistemic split. This means that subsemigroups $\mathcal{A}_{i}\subset
\mathcal{A}$ of the semigroup of observables of the \textit{unus mundus} are
established and identified, which correspond to conscious individuals and will
have a nonvanishing intersection with $\mathcal{A}_{\text{\textit{mind}}}$ of
eq \ref{matmind}. Moreover, the relationship between $\mathcal{A}_{i}$ and
$\mathcal{A}_{matter}$ will be largely a complementary one.

\item The \textit{unus mundus }itself is timeless, but after the epistemic
split, observables $T_{i}\in\mathcal{A}_{i}$ will be identifiable, which,
similar to the situation for the Wheeler- de Witt equation, due to the
entangledness of the state of the \textit{unus mundus}, assume time character,
monitoring other observables via entanglement correlations. Our mode of
personal existence reveals that $T_{i}$ will have the quality of an A-time in
the sense of ref \cite{mctaggart}. The quality of A-time will depend on the
level of consciousness. For simple organisms the notion of ''now'' will be the
predominant feature, and a faint notion of past will be able to incorporate
the results of learning from the environment. At higher levels, the notion of
past will be more elaborate, and a self model \cite{metzinger} will allow
planning of actions and the development of a differentiated notion of future.
So, the spectrum $specT_{i}$ will contain at least an element ''now'' and,
depending on the level of consciousness a simple or elaborate set of labels
pertaining to the more or less remote past and future. It is only by
entanglement correlations that $T_{i}$ assumes the quality of a time. Unlike
the situation for the Wheeler-de Witt equation we do not expect any strict
property of hyperbolicity to hold, because this would lead to a deterministic
dependence on $T_{i}$, which is highly implausible at this level of generality
and for the primarily individual A-time $T_{i}$.

\item For well separated different individuals we can expect their time
observables to commute:%
\begin{equation}
T_{i}T_{j}=T_{j}T_{i}%
\end{equation}
There will be entanglement correlations between different time observables
$T_{i}$ and $T_{j}$ giving a rough synchronization between them.

\item Entanglement correlations will also exist with material systems. These
correlations will be particularly strong for ''clock-like'' systems, for
instance certain astronomical systems. The observable semigroup $\mathcal{A}%
_{I}$ of these systems will contain clock observables $T_{I}$ which show
particularly strong entanglement correlations among each others and with the
variables $T_{i}$. Again, we expect commutativity%
\begin{equation}
T_{I}T_{J}=T_{J}T_{I},\;\;T_{i}T_{I}=T_{I}T_{i}%
\end{equation}
These strong correlations make it possible to transport a notion of time into
material systems and to attribute the quality of time also to the variables
$T_{I}$. However, the A-character of time will get lost in this transport
operation, and $T_{I}$ will rather look like a B-time.

\item Such processes of transportation and identification can be used to
construct a more and more universal and operationalized time by taking into
account more and more entanglement correlations and by choosing and redefining
time observables such as to maximize their entanglement correlations. This
process of purification and operationalization is really what happened in the
development of the notion of time in human thinking in general and in
particular in the development of science, eventually leading to the concept of
time in contemporary physics. With respect to this time, physical determinism
will hold, at least to a very good approximation and with respect to
$\mathcal{A}_{matter}$. Even the notion of internal time is reconsidered and
modified under the influence of physical time. This process leading to a clear
and sharp notion of B-time, of course required human consciousness at its
highest level. The same holds for a manifold of elaborations of B-time like
cyclic time or mythological time, which have been developed in various human societies.

The redefinition of an observable can easily be formalized in the framework of
Weak Quantum Theory: Let $A$ and $B$ be observables and take a function%
\begin{equation}
f:specA\longrightarrow specB
\end{equation}
Then we say that $B=f(A)$ if the following relations hold for the associated
projectors $A_{\alpha}$ and $B_{\beta}$ of eq \ref{spec}:%
\begin{equation}
B_{\beta}=\bigvee_{\alpha\in specA,\;f(\alpha)=\beta}A_{\alpha}%
\end{equation}
Just like in the previous chapter, the fact that physical time is normally
experienced as a classical quantity with a sharp value, is explained by a
decoherence mechanism. As compared to ordinary quantum theory, the situation
might be more favorable in Weak Quantum Theory, because, due to the absence of
a probability interpretation, the notion of a collapse of states is not
necessarily present in Weak Quantum Theory.

\item In the course of generalization and objectivation, time loses more and
more of its original qualities as A-time. Some steps on this way are: Internal
A-time, directed B-time and undirected B-time of time reversal invariant
physics. In contemporary physics, this process has even gone further. In parts
of string theory, as well as in quantum cosmology, timeless equations like the
Wheeler-de Witt equation have been formulated in which time has disappeared
altogether. Using a term employed by E. Ruhnau \cite{ruhnau} in a rather
different context, one might talk about \textit{deconstruction of time} as one
of the effects of the collective effort towards an increasing sharpening and
purification of the notion of time.
\end{enumerate}

Here, there may be the right place for a brief comparison of our approach with
the beautiful work of H. Primas \cite{primas2003}:

Primas tentatively applies ordinary quantum theory to the \textit{unus
mundus}. A first symmetry breaking leads to the decomposition into
(collective) mind and matter of eq \ref{primdec}. As opposed to our approach,
matter and mind observables are commuting rather than complementary. Time has
its origin in a one parameter symmetry of the timeless unus mundus and, after
the decomposition into matter and mind, appears with the representation of the
symmetry group in the collective mind sector. Sychronization with and transfer
to the matter sector is achieved by entanglement correlations which are a
consequence of the original symmetry of the unus mundus. It is reassuring and
it adds to the cogency of this picture to see the importance of entanglement
correlations also from a rather different approach. After the decomposition of
the \textit{unus mundus }into the commuting mind and matter sectors, the
scenario of Primas has much in common with Leibniz' view of a world governed
by prestabilized harmony, whereas we keep closer to the picture of Jung and
Pauli \cite{paulijung}, \cite{jungpaulibriefe}. A further difference is that
in his scheme B-time is born readily made in one step, while we try to
investigate the process of its stepwise emergence. Using full quantum
mechanics and representation theory of groups in Hilbert spaces, Primas
manages to derive a large number of interesting results and notions relevant
for the concept of time. He makes important remarks about the origin of the
directedness of time, which for us is present from the beginning, and about
the synchronization of the time arrows, even for non interacting systems, by
entanglement. In describing features of time in the mind sector he uses the
notion of a forward expanding Hilbert-space K-structure, which describes
learning and the filling up of a memory storage by the accumulation of
experience. Once time has been established along the route described above in
points 1 till 6, the related notion of an increasing sequence of propositions
can easily be incorporated into Weak Quantum Theory. A family of propositions
$(P_{\tau})_{\tau\in\mathbf{R}}$ can be called increasing, if%
\begin{equation}
P_{\tau}P_{\sigma}=P_{\sigma}P_{\tau}=P_{\sigma}\text{ for }\sigma\leq\tau
\end{equation}

\section{Questions, Observations, Speculations}

The issues raised in this last chapter are placed here not because we consider
them less important but because they lie somewhat off the main line of our argument.

\begin{itemize}
\item First of all, one should not forget that, in spite of its pervading
importance, the aspect of time cannot be applied to everything. On the
contrary, there are many observables, for instance observables pertaining to
logical questions or to issues of sense and meaning, which are unrelated or
complementary to time. There will be many observables $A$ with%
\begin{equation}
AT_{i}\neq T_{i}A
\end{equation}

\item Energy is a particularly clear and important example of such
observables. In ordinary quantum theory, the energy operator is conjugate to
time and generates time translations. The operator for a translation of time
by an amount $\alpha$ is given by%
\begin{equation}
U_{\alpha}=e^{2\pi i\alpha H/\text{$h$}}%
\end{equation}
where $H$ is the energy operator. The question now arises, whether the concept
of energy can be generalized in a qualified way such that it applies beyond
the realm of ordinary physics. The wider applicability of notions like
complementarity and entanglement has been demonstrated in Weak Quantum Theory.
The notion of time is never restricted to the domain of physics, and this
study was devoted to it. It would be desirable also to provide a qualified
generalization of the notion of energy. For instance, given a sufficiently
universal but not entirely physical time observable $T$, one would like to
define an operator $U_{\alpha}$ fulfilling a relation like%
\begin{equation}
U_{\alpha}TU_{\alpha}^{-1}=T+\alpha\label{trans}%
\end{equation}
There is, of course, an intuitive notion of energy used in everyday language,
and the notion of energy in physics arose from it by a process of sharpening
and operationalization similar to the one described for time above. In a very
vague sense, this notion is related to the capability of effectuating changes.
Associated to the intuitive notion of time is an element of will and desire,
which is one of the features which have got lost in the operationalization to
the energy in physics. A version of equation \ref{trans} may be able to
capture some features of the intuitive notion of energy. Quite generally,
energy should be related to any kind of transition. Transitions are
topicalized in \textit{process philosophy}. Normally, descriptions of the
functioning of human mind are centered on a discussion of its concepts,
notions and categories which can be associated to more or less stable states
of the mind. Emphasis may be shifted to transitions, which are just the
contrary of categorizations and genuinely acategorical \cite{atmanspacher}.
The generalized energy observable should be closely related to such
acategorical features of the human mind. This mental aspect of energy must not
be completely disjoint from its material side. In fact, Bekenstein
\cite{bekenstein} argues, that every exchange of information is associated to
a, however tiny, minimal exchange of energy.

\item We already emphasized on several occasions the paramount importance of
the epistemic split for every act of cognition. The very notion of an
observable already presupposes it, and the semigroup of observables is subject
to perspectiveness and depends on the observer. Now, given that the observer
is a conscious individual, and that time is intimately related to the form of
existence of conscious individuals, it would not be surprising to find
temporal features in any semigroup of observables. This is indeed the case.
The notion of composition of observables contains an embryonic element of time
in as far as $AB$ means ''$A$ applied \textbf{after} $B$ '', were ''after'' is
always meant in a temporal sense.

\item Time also enters in another way into the semigroup $\mathcal{A}$ of
observables. The state of the observer will change, not the least as a result
of the observations he makes. The observer dependence of the semigroup of
observables will thus render it time dependent, too. This change may result in
adding or modifying observables and also in the transformation of
preobservables, as described at the end of Chapter 2, into full fledged observables.

\item Finally, having discussed time at considerable length, one might wonder
about space. We expect also space to arise only after the epistemic split. As
opposed to time, it will have its origin in the material component
$\mathcal{A}_{matter}$ of the unus mundus. This corresponds well with
Descartes' attribution of space to the \textit{res extensae} and with the way
Kant interprets time as the form of the outer rather than the inner sense.
These questions certainly deserve a study of their own.
\end{itemize}

\textbf{Acknowledgement}

I should like to thank Harald Atmanspacher for many inspiring critical
discussions on the subject of this article and for valuable improvement
suggestions. Klaus Jacobi's philosophical insight was of great help for me.
Very special thanks are due to Georg Ernst Jacoby for inspiration and
continuous encouragement.

\end{document}